\documentclass[twocolumn,showpacs,preprintnumbers,amsmath,amssymb]{revtex4}
\usepackage{graphicx}
\usepackage{dcolumn}
\usepackage{bm}
\usepackage{here}

\begin{document}
\draft
\title{Rugged Fitness Landscapes of Kauffman Models with a Scale-Free Network}

\author{Kazumoto Iguchi\cite{byline1}, Shuichi Kinoshita\cite{byline2} and Hiroaki Yamada\cite{byline3}}

\address{{\it 70-3 Shinhari, Hari, Anan, Tokushima 774-0003, Japan\/}\cite{byline1}}
\address{{\it Graduate School of Science and Technology, Niigata University,\\
Ikarashi 2-Nochou 8050, Niigata 950-2181, Japan\/}\cite{byline2}}
\address{{\it YPRL, 5-7-14-205 Aoyama, Niigata 950-2002, Japan\/}\cite{byline3}}

\date{\today}


\begin{abstract}
We study the nature of fitness landscapes of 'quenched' Kauffman's 
Boolean model with a scale-free network.
We have numerically calculated 
the rugged fitness landscapes, 
the distributions,
its tails, and 
the correlation between the fitness of local optima 
and their Hamming distance from the highest optimum found,
respectively.
We have found that
(a) there is an interesting difference between
the random and the scale-free networks such that
the statistics of the rugged fitness landscapes is Gaussian
for the random network while it is non-Gaussian
with a tail for the scale-free network;
(b) as the average degree $\langle k \rangle$ increases,
there is a phase transition at the critical value of 
$\langle k \rangle = \langle k \rangle_{c} = 2$,
below which there is a global order and above which
the order goes away.
\end{abstract}

\maketitle



\section{Introduction}

The origin of life is one of the most important unsolved problems in science\cite{Maddox}.
To answer the quest, the self-organization of matter\cite{Eigen}
and the emergence of order\cite{KauffmanBook1} 
have been regarded as the key ideas.
To investigate such ideas, as early as in 1969 
Kauffman has introduced the so-called {\it $NK$ Kauffman model}
-- a random Boolean network model
based upon the random network theory\cite{ErdosRenyi}.
This model has been a prototype model
and studied by many authors
for a long time in understanding complex systems 
such as metabolic stability and epigenesis, genetic regulatory networks, and
transcriptional networks\cite{KauffmanBook1} as well as 
general Boolean networks\cite{DerridaPomeau,BasParisi,Newman,AldanaCK}, 
neural networks\cite{Hopfield} and spin glasses\cite{Anderson}.

At nearly the end of 1990's 
new kinds of networks, called scale-free networks, have been discovered
from studying the growth of the internet geometry and topology\cite{Strogatz,Barabasi,DM}.
After the discovery, scientists have known that
many systems such as those which were originally studied by Kauffman
as well as other various systems such as 
internet topology,
human sexual relationship, 
scientific collaboration,
economical network, etc.
belong to the category of the scale-free networks.
Therefore, it is very interesting for us to know what will happen when we apply
the concepts of scale-free networks to the Kauffman's Boolean network models.

Recently, there have just started some studies in this direction
\cite{LuqueSole,AlBarabasi,FoxHill,OosawaSava,Wang,Aldana,SerraVA,CastroSM}.
There, the Boolean dynamics of the Kauffman model with a scale-free network
has been intensively studied.
We would like to shortly call this the scale-free ($SF$) Kauffman model.
Hence, there are many interesting problems that are necessary to be considered.

Thus, in this paper we would like to study the structure of rugged
fitness landscapes of the $SF$ Kauffman model.
We would like to know what is the difference in fitness landscapes between
the $NK$ and the $SF$ Kauffman models.

The organization of the paper is the following:
In Sec.II, we summarize the formalism on Boolean dynamics of
the $NK$ Kauffman model.
In Sec.III, we introduce the formalism to calculate the statistics and
the rugged fitness landscapes of $NK$ Kauffman model.
And our scheme to obtain a scale-free network is given.
In Sec.IV, we show the numerical results of the rugged fitness landscapes,
of the histograms and its tails,
and of the correlation between the fitness of local optima 
and their Hamming distance from the highest optimum found,
for both 'quenched' $NK$ and $SF$ models, respectively.
In Sec.V, our conclusion will be given.


\section{$NK$ Kauffman model}

In the $NK$ model (the $SF$ model is described below), we assume that 
the total number of nodes (vertices) $N$ and 
the degree (i.e., the number of inputs) of the $i$-th node $k_{i}$ 
in the network are fixed such that all $k_{i} = K$.
Therefore, the resultant graph is a {\it directed} random network,
where each link has its own direction as represented by an arrow on the link. 
This gives us in general an {\it asymmetric} adjacency matrix of network theory. 
Since there are $K$ inputs to each node, 
$2^{K}$ Boolean spin configurations can be defined on each node;
the number $2^{K}$ certainly becomes very large as $K$ becomes a large number.

We then assume that Boolean functions are randomly chosen on each node
from the $2^{2^{K}}$ possibilities.
Locally this can be represented by
$$\sigma_{i}^{(t+1)} = B_{i}(\sigma_{i}^{(t)};\sigma_{i_{1}}^{(t)},
 \cdots, \sigma_{i_{k_{i}}}^{(t)}), \eqno{(1)}$$
for $i = 1, \dots, N$,
where $\sigma_{i}  \in Z_{2} \equiv \{0,1\}$ is the binary state and
$B_{i} \in Z_{2}$ is a Boolean function at the $i$th node, 
randomly chosen from $2^{2^{k_{i}+1}}$ Boolean functions with
the probability $p$ (or $1-p$) to take 1 (or 0).

If we fixed the set of the randomly chosen Boolean functions 
$\{B_{i}, i =1, \cdots, N \}$
in the course of the time development,
then this model is called the {\it quenched model}\cite{KauffmanBook1}.
On the other hand, 
if we change the set each time, then this model is called the {\it annealed model}
\cite{DerridaPomeau,BasParisi}.
If we study the dynamics of the states in the system taking care of Eq.(1),
then we are able to obtain the cyclic structures of the states such as
the length of the cycle, 
the transient time and 
the basin sizes, etc. 
These are usually calculated numerically, 
since it is extremely difficult to do the calculations
analytically\cite{KauffmanBook1,Newman,AldanaCK}.

However, in the annealed models\cite{DerridaPomeau,BasParisi},
it has been investigated analytically that
as the degree of nodes $K$ is increasing,
there exists a kind of {\it phase transition of network}
at the critical degree $K_{c} = 1/[2p(1-p)]$
and if we conversely solve it for $p$ then
we obtain the critical probability $p_{c} = (1 \pm \sqrt{1-2/K})/2$.


\section{Fitness Landscape model}

Let us now study {\it statistics} in the structure of the fitness landscapes 
of the $NK$ and $SF$ Kauffman models.
The fitness landscapes are calculated as follows\cite{KauffmanBook1}:
(i) Generate a network with $N$ nodes and the degree $k_{i}$ of the $i$-th node.
The $k_{i}$ links are inputs that are directed to the $i$-th node.
(ii) Define the local fitness at the $i$-th node by
$$W_{i} = f_{i}(\sigma_{i};\sigma_{i_{1}}, \cdots, \sigma_{i_{k_{i}}}) \eqno{(2)}$$
for $i = 1, \dots, N$,
where $W_{i}$ takes one of $2^{k_{i}+1}$ real numbers $\omega_i$
which are randomly taken from the interval $[0,1]$.
This provides us a table for each node (TABLE.1).
(iii) Define an $N$-component initial state $A$, say $\Psi_{A} = (0, 1, 1, \dots, 0)$.
(iv) Investigate the input states on $k_{i}$ links for the $i$-th node.
And adjusting the states in the entries with the table,
choose the fitness $W_{i}$ from the $2^{k_{i}+1}$ values of 
$w_{1}, w_{2}, \cdots, w_{2^{k_{i}+1}}$.
Then, define the fitness $W_{A}$ for the state $A$ by
$$W_{A} = \frac{1}{N} \sum_{i=1}^{N} W_{i}.                         \eqno{(3)}$$
(v) Each state forms a vertex of the $N$-dimensional hypercube
so that there are totally $2^{N}$ vertices,
each of which has its $N$ neighbors
such as $\Psi_{B} = (1, 1, 1, \dots, 0)$
(i.e., one-mutant variants, denoted by $n_{omv}$).
Then, calculate the fitnesses $W_{B}$ for these neighbor states
in the same way.
(vi) Compare the fitness value $W_{A}$ of the $A$ state
with those of the neighbor states such as $W_{B}$, successively.
Here the Hamming distances between the state and its neighbors are all $1$.
If $W_{A} > W_{B}$ for all neighbors $B$'s, then
the fitness $W_{A}$ for the $A$ state is a local optimum. 
And if we meet a neighbor $B$ such that $W_{A} < W_{B}$,
then write $B = A'$ and $W_{B} = W_{A'}$.
Redo the same procedure 
with all neighbors to obtain $W_{A''}$, $W_{A'''}$, etc. 
until the local optimum is found.
(vii) Finally, measure the difference between
each fitness of the neighbors and the local optimum fitness.
This provides us a rugged fitness landscape of the system.

The above procedure starts from the particular initial state
with the set of the random numbers $W_{i}$. 
Since we can change either the initial state to a different state in the $2^{N}$ states
or the set to a different set chosen randomly,
we can generate many samples.
Each sample results in a different rugged fitness landscape of the system.
Hence, we obtain an ensemble of them.
Thus we can study the statistics of the structures of rugged fitness landscapes\cite{Note1}.
In this paper we take a thousand samples for the purpose.
\\
\newline
\begin{tabular}{|ccccc|c|}  \hline \hline
{\em $\sigma_{i_{1}}$} & {\em $\sigma_{i_{2}}$} & {\em $\cdots$} 
& {\em $\sigma_{i_{k_{i}}}$} & {\em $\sigma_{i}$}& 
{\em $W_{i}$}\\  \hline 
$0$&$0$&$\cdots$&$0$&$0$&$\omega_{1}$\\ 
$0$&$0$&$\cdots$&$0$&$1$&$\omega_{2}$\\
$0$&$0$&$\cdots$&$1$&$0$&$\omega_{3}$\\ 
$0$&$0$&$\cdots$&$1$&$1$&$\omega_{4}$\\ 
$\vdots$&$\vdots$&$\vdots$&$\vdots$&$\vdots$&$\vdots$\\
$1$&$1$&$\cdots$&$1$&$1$&$\omega_{2^{k_{i}+1}}$\\ \hline\hline
\end{tabular}\\
\\
TABLE 1.  The relationship between the (real) output $W_{i}$
and the (binary) inputs, $\{\sigma_{i_{1}}, \sigma_{i_{2}}, \cdots,
\sigma_{i_{k_{i}}}, \sigma_{i}\}$.
Since there are ($k_{i}+1$) $\sigma_{i}$'s,
each of which has $0$ or $1$,
there are $2^{k_{i}+1}$ ways of inputs.
These provide $2^{k_{i}+1}$$\omega_{i}$'s,
each of which is a real number randomly drawn from the interval $[0, 1]$,
according to a homogeneous distribution
\\



\begin{figure}[h]
\includegraphics[scale=0.55]{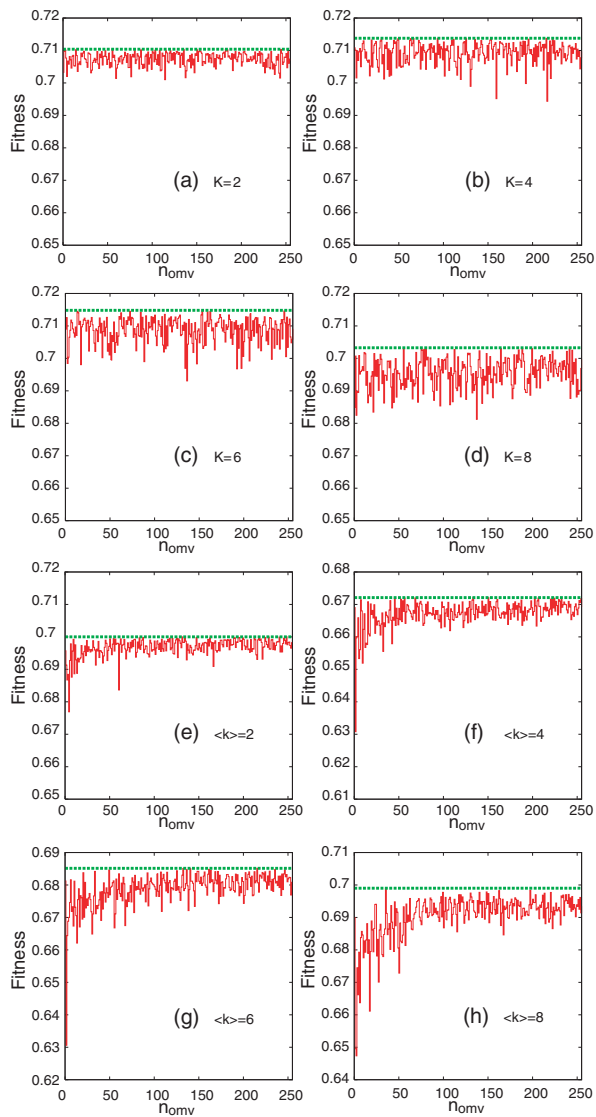}
\caption{
(Color online)
Rugged fitness landscapes of the 'quenched' random $NK$ 
and of the 'quenched' $SF$ Kauffman models,
where the total number of nodes is given by $N = 256$.
(a)--(d) are shown for the random networks of $K = 2, 4, 8, 16$, respectively,
and (e)--(h) for the scale-free networks
of $\langle k \rangle = 2, 4, 8, 16$, respectively,
where $\langle k \rangle$ means the average degree of nodes.
In each figure, the vertical axis shows the fitness 
while the horizontal axis shows the family of 
all one-mutant variants $n_{omv}$ of a local optimum.
The dotted line denotes the fitness level of the local optimum
and 
the curve the fitness differences between
the local optimum and all the one-mutant variants.
} 
\label{fig:1}
\end{figure}

To apply the above method to the $SF$ Kauffman model,
we have to specify a model for the scale-free network
with an arbitrary degree of nodes $\langle k \rangle = n$, even integer.
For this purpose,
let us adopt a slightly modified version of 
the so-called Albert-Barab\'{a}si (AB) model\cite{Barabasi}
as a prototype model.
In our model,
we initially start with $m_{0} = n/2$ nodes for seeds of the system, 
all of which are linked to each other such that the total link number is $n(n-2)/8$.
And every time when we add one node to the system, 
$m = n/2$ new links are randomly chosen in the previously existing network, 
according to the preferential attachment probability of 
$\Pi_{i}(k_{i})=k_{i}/\sum_{j=1}^{N-1}k_{j}$.
Then, after $t$ steps, we obtain
the total numbers of nodes $N(t) = n/2 + t$ and 
of links $L(t) = n(n-2)/8 + (n/2)t$, respectively.
We continue this process until the system size $N$ is achieved.
Hence, by this we can define 
$\langle k \rangle \equiv 2L(t)/N(t) = n$ as $t \rightarrow \infty$.
The generalization can be straightforward.
Now we apply the above-mentioned
dynamics to this modified AB model and call the result the $SF$ Kauffmann
model \cite{KauffmanBook1}.

\begin{figure}[h]
\includegraphics[scale=0.45]{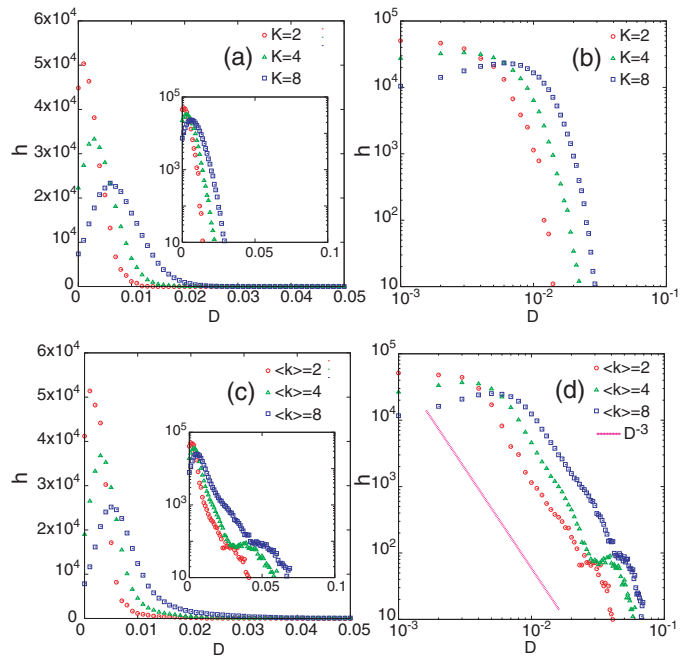}
\caption{
(Color online)
The number of times $h$ when a particular value of the fitness difference $D$
between a local optimum and $n_{omv}$ occurs in the rugged fitness landscapes.
Here the 'quenched' models have been adopted.
(c) and (d) [(a) and (b)] show the histograms and the tails 
for the scale-free [random] networks of
$\langle k \rangle = (K =) 2$ ($\bigcirc$), $4$ ($\Box$), $8$ ($\triangle$), 
respectively, where $N =256$.
Here $\langle k \rangle$ stands for the average degree of nodes.
The data for 1000 samples are superimposed in each figure.
(b) and (d) (inset) are shown in the log-log (semi-log) plot of the tails.
} 
\end{figure}


\section{Numerical results}
\subsection{Rugged fitness landscapes}

Fig.1 shows the rugged fitness landscapes of
both the 'quenched' random $NK$ and the 'quenched' $SF$ Kauffman models.
Here we have calculated the systems up to $N = 256$.
This is just because of our computer power at this moment.
We only show the data for $N = 256$ in this paper.
As previously noted by Kauffman\cite{KauffmanBook1}
the fitness landscapes of the random networks
are very rugged.
We see that the fitness landscapes of scale-free networks
are very rugged as well, 
but quite different from those of the random networks. 
This can be understood as follows:
In the random network each one of the nodes
always meets with the same $K$ links to the neighbors.
The ruggedness can be dominantly bounded by the value of $K$.
Hence, as $K$ becomes large, the fluctuations in the
rugged fitness landscapes become large.
On the other hand,
in the scale-free network there are various kinds
of degrees $k_{i}$ of nodes.
In other words, each node has its own $k_{i}$ links
and there is the distribution of the degrees
such as $P(k) \propto k^{-\gamma}$.
The AB model exhibits power-law with $\gamma = 3$.
Therefore, the fitness landscapes can fluctuate,
according to the degree distribution of the network.
So, as the degree $k_{i}$ of a node is large, the difference in the fitness
between the optimum and the mutants is expected to be large.
Hence, the fitness landscapes obey the nature of the scale-free network.


\subsection{Histograms and tails of rugged fitness landscapes}

How can we detect the differences in the rugged fitness landscapes 
between the random and the scale-free networks?
To do so, 
denote by $h$ the histogram and denote by $D$
the fitness difference between the local optimum and the one-mutant variant.
Then we draw the histograms of the rugged fitness landscapes 
in the normal plot [(a) and (c)],
in the log-log plot [(b) and (d)],
and in the semi-log plot (insets), respectively (Fig.2).
Comparing (a) with (c) in the numerical results, 
we find that
the histograms of the fitness landscapes for the random networks
behave like Gaussian distributions, which can be fitted by
$h \propto e^{-(D - \langle D\rangle)^{2}/v^{2}}$
where we set the peak value as $\langle D \rangle$, 
the variance as 
$v = \sqrt{\langle (\Delta D)^{2}\rangle}$
with $\Delta D = D - \langle D \rangle$.
We have found numerically that
$\langle D \rangle = \{0.943,2.62, 5.85\}\times 10^{-3}$;
$v = \{4.31, 5.75, 7.58\}\times 10^{-3}$
for $K = 2, 4, 8$, respectively.
On the other hand, 
the histograms for the scale-free networks
behave like non-Gaussian distributions with a broad tail,
which can be fitted by
$$h \propto 
\left\{ 
\begin{array}{ll}
e^{-\frac{(D - \langle D \rangle)^{2}}{v^{2}}} & \mbox{if $(D-\langle D \rangle)^2 \ll  v^2$} \\
e^{-\alpha \frac{D - \langle D \rangle}{v}} & \mbox{if $D$ is comparable to  $\langle D \rangle$}\\
D^{-\beta} & \mbox{if $D$ is much larger than $\langle D \rangle$}
\end{array}
\right..  
\eqno{(4)}$$
Here we have found numerically that
$\langle D \rangle = \{1.48, 2.74, 5.27\} \times 10^{-3}$;
$v = \{3.45, 4.67, 5.53\} \times 10^{-3}$;
$\alpha = 1.30, 1.21, 0.978$;
$\beta = 3.11, 3.50, 2.98$,
for $\langle k \rangle = 2, 4, 8$, respectively.

We note here the following: (a) The tail (i.e., scaling behavior) appears
when the system size becomes as large as $N = 256$. And as the value of
$\langle k \rangle$ is increasing, the value of $\beta$ seems closer to
$\gamma = 3$ of $P(k)$ in the AB model\cite{Barabasi} [Fig.2 (d)]. But in
view of the limited accuracy of Fig. 2(d) (e.g., $\beta = 3.50, 2.98$), at
the present moment this can be a {\it conjecture}, speculated from the
numerical results for the system of $N = 256$.  However, in Appendix A we
give a sharp analytical arguments showing a strict relationship between the
power-law decay and the "degree" fluctuations.  And also, since we extend
the system up to $N = 1024$, we are able to confirm ourselves that the tail
behavior is maintained and become more prominent, as $N$ is increasing. As
an example, we show the result in Appendix B.

(b) The results for the random networks
show a kind of {\it transition} when the value of $K$ is
going up from $K = 2$ to $K = 8$.
This is consistent with the critical value of $K_{c} = 2$ for $p = 0.5$
which was analytically obtained from the annealed model\cite{DerridaPomeau}.
Therefore, the distribution below $K_{c}$ is quite different from that above $K_{c}$
so that the distributions for $K > K_{c}$ become more Gaussian-like as $K$ increases.
Very interestingly, we find a similar transition 
for the scale-free networks as well,
when the value of $\langle k \rangle$ is
going up from $\langle k \rangle = 2$ to $\langle k \rangle = 8$.

This can be explained as follows:
Suppose the distribution of degrees is approximately given by
$P(k) = k^{-\gamma}/\zeta(\gamma)$
such that we can impose normalization 
$\sum_{k=1}^{\infty} P(k) =1$,
where $\zeta(\gamma)$ is the Riemann's zeta function defined by
$\zeta(\gamma) = \sum_{k=1}^{\infty} k^{-\gamma}$.
Substituting it to the definition
$\langle k \rangle = \sum_{k=1}^{\infty}kP(k)$,
then we obtain
$$\langle k \rangle = \zeta(\gamma-1)/\zeta(\gamma),    \eqno{(5)}$$
which is finite for $\gamma > 2$ and infinite for $ 1 < \gamma < 2$
and which was first obtained by Aldana et. al.\cite{Aldana}.
For example, since $\gamma = 3$ for the special case of the AB model, 
we obtain
$\langle k \rangle = \zeta(2)/\zeta(3) = 1.64493\cdots/1.20205\cdots \approx 1.3684$.
As studied by Aldana et. al.\cite{AldanaCK,Aldana},
the critical value $\langle k \rangle_{c}$  for the annealed dynamics of the $SF$ Kauffman model
is given by 
$\langle k \rangle_{c} = \zeta(\gamma_{c}-1)/\zeta(\gamma_{c}) = 2$ for $p = 0.5$
as well, where $\gamma_{c} \approx 2.47875$.
Within the limited accuracy of our numerical
results (some 20 percent) this is the same value as just stated for the
quenched dynamics. 
In fact, we expect that also for our quenched dynamics there is a critical
point around  $\langle k \rangle_c=2$ , with $\beta$ around 3, maybe again
exactly at these values. In fact, in appendix A we show that also in our
case the statistics of the rugged fitness landscape is bounded by $K$, such
that  $\beta$ reflects the fluctuations of the 'degree', which should be the
same both for the quenched and the annealed dynamics
\cite{AldanaCK,Aldana}.
Thus, our numerical results support this analytical result
although our system is not very large but a finite scale-free network of $N = 256$.
We also note here that we confirm that
the phase transition at the critical value of
$\langle k \rangle = \langle k \rangle_{c} = 2$,
which was proved analytically in the 'annealed' model
by Aldana {\it et.al.} \cite{Aldana},
occurs in the 'quenched' model as well \cite{IKY}.


\subsection{The correlation between the fitness of local optima 
and their Hamming distance from the highest optimum found}

\begin{figure}[h]
\includegraphics[scale=0.6]{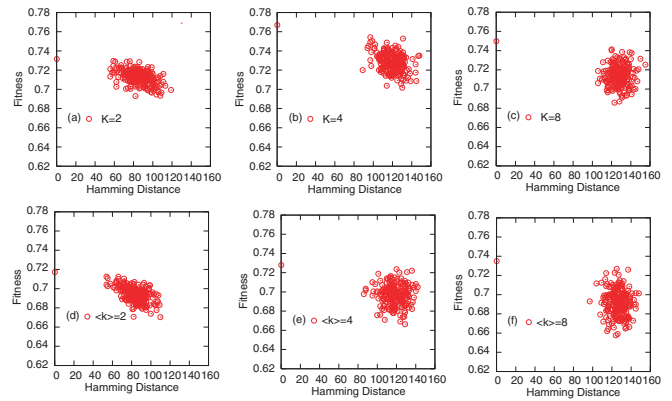}
\caption{
(Color online)
The correlation between the fitness of local optima 
and their Hamming distance from the highest optimum found.
It is shown for the random networks with $K = 2$ (a), $4$ (b), $8$ (c)
and for the scale-free networks with $\langle k \rangle = 2$ (d), $4$ (e), $8$ (f),
respectively, where $N = 256$.
The vertical axis stands for the fitness and 
the horizontal axis the Hamming distances 
between the largest local optimum and the local optima.
The data for 1000 samples are superimposed in each figure.
In both cases there seem to exist phase transitions,
i.e., at  $K = K_{c} = 2$  for the random networks
and $\langle k \rangle = \langle k \rangle_{c} = 2$ for the scale-free networks,
supporting the exact statement for the 'annealed' version of the
first mentioned case and the analytical prediction of Aldana et. al. on
the annealed version of the $SF$ model.
} 
\end{figure}

Finally we present the correlation between the fitness of local optima 
and their Hamming distance from the highest optimum found\cite{KauffmanBook1} (Fig.3).
In both the random and the scale-free networks we find the following: 
If $K$ and $\langle k \rangle$ are as small as the critical value of $2$,
then the highest optima are nearest to one another. 
And the optima at successively greater Hamming distances from the
highest optimum are successively less fit.
Therefore, there is a global order to the landscape.
On the other hand,
as $K$ and $\langle k \rangle$ increase, the correlations fall away.
This shows that the previous assertions are maintained
in the correlations, respectively.

\section{Conclusions}

In conclusion, we have studied the structure and statistics of the
rugged fitness landscapes for the quenched $SF$ Kauffman models,
comparing with that for the quenched random $NK$ Kauffman models.
We have numerically calculated the rugged fitness landscapes, the distributions,
the tails, and the fitness correlations of local optima with the Hamming distance
from the highest optimum, respectively.
From the results, we have concluded that 
in the $SF$ Kauffman models there is a transition of network 
when $\langle k \rangle = \langle k \rangle_{c} = 2$,
while in the $NK$ Kauffman models such a transition occurs
at $K = K_{c} = 2$.
This is, in some sense, quite analogous to the situation in the
study of Boolean dynamics of $NK$ and $SF$ Kauffman models\cite{AldanaCK}.
It would be very interesting if we could apply this approach to 
study fitness landscapes of other network systems.

\acknowledgments
We would like to thank Dr. Jun Hidaka for collecting many relevant papers.
One of us (K. I.) would like to thank Kazuko Iguchi for her continuous
financial support and encouragement.

\appendix
\section{Analytical arguments of relation between the power-law decay and the 'degree'
fluctuations}
The tail behavior of the histogram is interpreted as follows.

Define a network with $N$ nodes,
which can be any kind of network such as random, scale-free and exponential-fluctuating networks,
where the $i$-th node has $k_{i}$ degree (i.e., link number).
Let us consider the Kauffman model.
As given in Table 1, the number of inputs to the $i$-th node 
is given by $k_{i}+1$.
Let us now choose an $N$-dimensional initial state $A$
such as $(1, 0, 0, 1, 1, \cdots, 1)$.
Considering the input from the given network structure,
we can define fitness $W_{i}$ on each node.
Therefore, we can define the fitness $W_{A}$ of the state $A$
by 
$$W_{A} = \frac{1}{N} \sum_{j=1}^{N}W_{j}^{A}, \eqno{(A1)}$$
where $W_{j}^{A}$ means fitness at the $j$-th node in the state $A$.
Consider one-mutant family of the initial state $A$,
in which there are $N$ neighbor states with Hamming distance one.
For example, let us say one of them, $B$
and define as $(0, 0, 0, 1, 1, \cdots, 1)$.
In this example, only the state of node 1 is different from 1 to 0. 
Therefore, the difference in input values between the initial state $A$ 
and this state $B$ comes from nodes linked to node 1.
This situation provides a difference in fitness.

Suppose that the degree of node 1 is $k_{1}$ and 
denote the $k_{1}$ nodes linked to node 1 by $j_{1}, \cdots, j_{k_{1}}$.
The values of nodes linked to node 1 are given different random numbers $W_{i}'$
according to Table 1.
We then obtain fitness for the state $B$ as
$$W_{B} = \frac{1}{N} \sum_{j=1}^{N}W_{j}^{B}
= \frac{1}{N}(W_{1}' + W_{j_{1}}' + \cdots + W_{j_{k_{1}}}' + \cdots).\eqno{(A2)}$$
Thus, a genetic mutation in the state of one-mutant gives
a change only for the node that the mutation occurred and the nodes linked to it.
Therefore, if we consider only the fitness difference from the local optimum,
then the fitness value of the one-mutant family that has Hamming distance 1 from
the local optimum state depends upon which node the mutation occurs.
Hence, in the case that there is mutation on node $m$,
we obtain
$$\Delta W_{B}(m)
= \frac{1}{N}(\Delta W_{1} + \Delta W_{j_{1}} + \cdots + \Delta W_{j_{k_{1}}}),\eqno{(A3)}$$
where $\Delta W_{j} = W_{j}' - W_{j}$.
The left hand side of Eq.(A3) means the fitness difference, $D$.
To see what it means, let us define
the averaged fitness difference for node $m$:
$$\langle D_{B}(m) \rangle 
= \frac{1}{k_{m}+1}(\Delta W_{1} + \Delta W_{j_{1}} + \cdots + \Delta W_{j_{k_{m}}}). \eqno{(A4)}$$
We then have
$$D_{m} = \Delta W_{B}(m) = \frac{1}{N} \langle D_{B}(m) \rangle (k_{m}+1). \eqno{(A5)}$$
From this, if the average $\langle D_{B}(m) \rangle$ is constant,
then $D_{m}$ is proportional to $k_{m} + 1$.
But, more generally,
there are two contributions: one from random number $W_{i}$ in
$\Delta W_{1} + \Delta W_{j_{1}} + \cdots + \Delta W_{j_{k_{m}}}$
and another from degree $k_{m}$.
Here, if the random numbers are defined by a uniform distribution,
then we can understand that they contribute to the exponent 
of the fitness distribution function
and the tail of the distribution function comes from 
that of the degree (link number).
Because, since the maximum value of $W_{i}$ is 1,
it is bounded as
$$|\Delta W_{1} + \Delta W_{j_{1}} + \cdots + \Delta W_{j_{k_{m}}}| \le k_{m}+1.\eqno{(A6)}$$
Hence, this provides
$$|D_{m}| \le \frac{k_{m}+1}{N}. \eqno{(A7)}$$
From the above, in the $NK$ model
the statistics of rugged fitness landscapes is bounded by $K$.
In the $SF$ model, since $k_{i}$ is distributed by
a power law, the statistics becomes the same power distribution.
Similarly, in the exponential fluctuation distribution,
so is the fitness distribution.
In this way, the statistics of fitness is strongly
dominated by that of the link distribution in the network.

\section{Tail behavior of the system of $N = 1024$}
We show the tail behavior of the histogram in the system size of $N =1024$
in Fig.4. This may support our assertion in the text.
We find again $\gamma \approx 3$ ; but since here the power-law decay is already obtained for
$<k>=2$, we cannot exclude that$<k>_c < 2$, although such a statement
would perhaps only reflect finite-size effects.

\begin{figure}[h]
\includegraphics[scale=1.5]{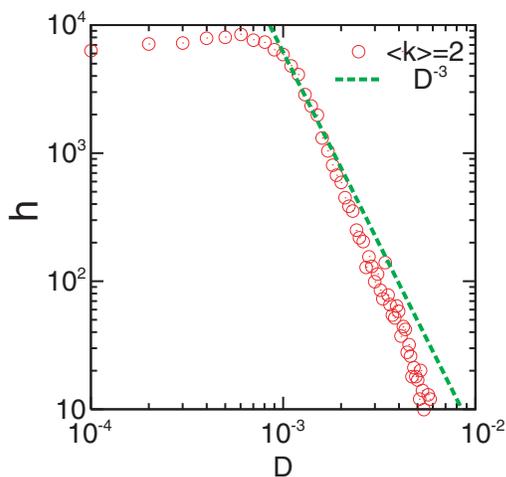}
\caption{
(Color online)
$log$-$log$ plot of the tail behavior of the histogram for scale-free networks
of the system size $N =1024$, where $\langle k \rangle = 2$.
Here the 'quenched' models have been adopted.
The vertical axis means $\log$ of the number of times $h$ 
when a particular value of the fitness difference $D$
between a local optimum and $n_{omv}$ occurs in the rugged fitness landscapes.
The horizontal axis means $log$ of the fitness difference $D$.
The data for 100 samples are superimposed in figure.
We confirm ourselves that $\gamma \approx 3$.
} 
\end{figure}

\end{document}